\documentclass[a4paper]{aa}
\usepackage{psfig}
\usepackage{epsfig}
\usepackage{graphicx}

\def \xmm {XMM-Newton}

\def \src {MXB\thinspace1659-298}
\def \degmark{^\circ}
\def \nh {$N{\rm _H}$}

\def \hcm {\hbox {\ifmmode $ atom cm$^{-2}\else atom cm$^{-2}$\fi}}

\def \arcsec {\hbox{$^{\prime\prime}$}}
\def \chisq {$\chi ^{2}$}

\def\approxgt{\mathrel{\hbox{\rlap{\lower.55ex \hbox {$\sim$}}
        \kern-.3em \raise.4ex \hbox{$>$}}}}
\def\approxlt{\mathrel{\hbox{\rlap{\lower.55ex \hbox {$\sim$}}
        \kern-.3em \raise.4ex \hbox{$<$}}}}
\newcommand{\mc}{\multicolumn}
\newcommand {\Msun}{M_\odot}
\begin{document}


\title{An XMM-Newton study of the
X-ray binary \src\ and the discovery of narrow X-ray absorption lines}

\author{L. Sidoli
        \and T. Oosterbroek 
        \and A. N. Parmar
        \and D. Lumb
        \and C. Erd
}
\offprints{L.Sidoli (lsidoli@astro.estec.esa.nl)}

\institute{Space Science Department of ESA, ESTEC,
        Postbus 299, 2200 AG Noordwijk, The Netherlands
}

\date{Received: 20 July 2001; Accepted: 21 September 2001 }

\authorrunning{L. Sidoli et al.}

\titlerunning{{\xmm\ observation of \src: discovery of X-ray absorption lines}}

\abstract{We report the discovery of narrow X-ray absorption lines from the
low-mass X-ray binary \src\ during an XMM-Newton observation in 
2001 February. The 7.1~hr orbital cycle is clearly evident
with narrow X-ray eclipses preceded by intense dipping activity.
A sinusoid-like OM $B$-band modulation with a peak-to-peak
modulation of $\sim$0.5 magnitude and a minimum coincident with the
X-ray eclipse is visible. 	
EPIC and RGS spectra reveal the presence of narrow 
resonant absorption features identified with O~{\sc viii}~1s-2p, 1s-3p and 
1s-4p,
Ne~{\sc x} 1s-2p, Fe~{\sc xxv}~1s-2p, and Fe~{\sc xxvi}~1s-2p transitions,
together with a broad Fe emission feature at $6.47 ^{+0.18} _{-0.14}$~keV. 
The equivalent widths of the Fe 
absorption features show no obvious dependence on orbital phase, 
even during dipping intervals. The equivalent widths of the other
features are consistent with having the same values during persistent
and dipping intervals. 
Previously, the only X-ray binaries known to exhibit narrow X-ray absorption 
lines were two superluminal jet sources and it had been suggested that
these features are related to the jet formation mechanism. This now
appears unlikely, and instead their presence may be
related to the viewing angle of the system.
The \src\ 0.6--12~keV continuum is modeled using 
absorbed cutoff power-law and blackbody components. During dips the
blackbody is more strongly absorbed than the power-law. 
The spectral shape of the 3.6\% of 0.5--10~keV emission that
remains during eclipses 
is consistent with that during non-dipping intervals.
} \maketitle
 
\section{Introduction}
\label{sect:intro}

\src\ is a transient bursting X-ray source discovered by SAS-3 (Lewin et
al. \cite{l:76}). Observations with
SAS-3 and HEAO-1 detected a clearly modulated persistent
emission which showed erratic intensity variations followed by 15
minute duration X-ray eclipses which repeated periodically (Cominsky
\& Wood \cite{cw:84}).  By combining data spanning $\sim$2~years an
orbital period of 7.11~h was obtained (Cominsky \& Wood \cite{cw:89}).
Following the discovery using EXOSAT of a second bursting, dipping, eclipsing,
transient (EXO\thinspace0748-676; Parmar et al. \cite{p:86}) the
region of sky containing \src\ was observed by EXOSAT. However, the
source was not present. Indeed, until recently \src\ has not been
observed to be active.  However, on 1999 April~2, after a
21 year quiescent interval, an 18 mCrab (2--9~keV) source at a
position consistent with \src\ was detected in one of the BeppoSAX
Wide Field Cameras (in 't Zand et al. \cite{i:99a}).  
Observations a
day later confirmed that the $V$=18.3 optical counterpart to \src\
(V2134~Oph; Doxsey et al. \cite{d:79}) had brightened significantly
and was exhibiting high excitation He~{\sc ii} emission, confirming
that the source was undergoing an outburst (Augusteijn et al.\
\cite{az:99}).

\src\ was observed by RXTE a few days later (Wachter et
al. \cite{w:00}) confirming that the source had resumed
its strong and persistent X-ray emission.  
Contemporaneous observations revealed that the optical emission
exhibits a 0.2 magnitude eclipse feature coincident with the X-ray
eclipse. The 0.5--30~keV BeppoSAX spectrum obtained in 2000 August
was modeled by a combination of a soft disk-blackbody and a harder
Comptonized component (Oosterbroek et al.~\cite{o:01}). The residuals
to this fit suggest the presence of emission features due to
Ne-K/Fe-L and Fe-K in the spectrum.  

Only two Low-Mass X-ray binary (LMXRB) systems are known that exhibit
both dips and eclipses (about 10 exhibit only dips).  
During
dips the observed spectral changes are   inconsistent with
simple absorption from cold material, as might be expected.  Detailed
modeling of these spectral changes provides a powerful
means of studying the structure and location of the emitting and
absorbing regions in LMXRB (e.g., Parmar et al. \cite{p:86}; Church \&
Ba\l uci\'nska-Church \cite{cb:95}; Church et al. \cite{c:97}). This
modeling has revealed that at least two emission components are
required which appear to undergo different amounts of 
absorption during
dips. In the case of EXO\thinspace0748-676, (the other eclipsing and
dipping LMXRB), recent XMM-Newton Reflection Grating Spectrometer
(RGS) observations have revealed the presence of discrete structures
due to absorption and emission from ionized Ne, O and N
(Cottam et al. \cite{c:01}). Simultaneous
European Photon Imaging Camera (EPIC) data revealed that the 0.2--10~keV
spectra can be fit with a two component model consisting of a central
Comptonized component and a more extended thermal halo
(Bonnet-Bidaud et al.\ \cite{bb:01}). The spectral variations observed
during dips were mainly accounted for by variations in the absorbing
column affecting the Comptonized component only. 

\begin{table}
\caption{\xmm\ instrument configurations}
\label{tab:log}
\begin{tabular}[c]{lccl}
\hline\noalign{\smallskip}
Instrument & Energy  	& \mc{1}{c}{Net}  & Mode \\
            & Range  &  Exposure	&  \\
            & \mc{1}{c}{(keV)}   & \mc{1}{c}{(ks)}      	&  \\
\noalign{\smallskip\hrule\smallskip}
EPIC PN 	&   0.2--15  	&  21.5      & Small window\\
EPIC MOS1&   0.2--15   	&  30.0      &       Timing \\
EPIC MOS2&   0.2--15   	&  30.0      &       Full window\\
RGS  	&   0.35--2.0  	&  31.1      & Spectroscopy + Q\\
OM       & \dots         & 20 $\times$~1 & B filter\\

\noalign{\smallskip\hrule\smallskip}
\end{tabular}
\end{table}

\section{Observations and Data Analysis}
\label{sect:obs}

The XMM-Newton Observatory (Jansen et al. \cite{j:01}) includes three
1500~cm$^2$ X-ray telescopes each with an EPIC 
at the focus. Behind two of the telescopes there are RGS arrays 
(den Herder et al. \cite{dh:01}). 
In addition, a coaligned optical/UV monitor telescope (OM,
Mason et al. \cite{m:01}) is included as part of the payload.
Two of the EPIC imaging spectrometers use MOS type CCDs (Turner et al.
\cite{t:01}) and one uses a PN type CCD (Str\"uder et al. \cite{s:01}).

The data analysis reported here used 
event lists produced by the standard \xmm\
SAS (Science Analysis Software, V.5.0.0) {\sc EMPROC}, {\sc EPPROC} and
{\sc RGSPROC} 
tasks which were further filtered using {\sc XMMSELECT} V.2.35.4. 
For the PN only pattern 0 (single pixel) events were selected.
Known hot, or flickering, pixels and electronic noise were rejected 
using the SAS and the low-energy cutoff was set to 0.2~keV. 
For the MOS, X-ray events corresponding to patterns 0--12 
were selected.

Source counts were extracted from circular regions of 40\arcsec\
radius centered on \src\ for the PN instrument.
Background counts were obtained from similar regions
offset from the source position. 
RGS source and background events were extracted by making spatial and
energy selections on the event files. 
Wavelengths were then assigned to the dispersion
coordinates using the latest calibration parameters.
In order to ensure applicability 
of the \chisq\ statistic, extracted spectra were
rebinned such that at least 20 counts per bin were present and such
that the energy resolution was not oversampled by more than a
factor 3. In order to account for systematic effects a 2\%
uncertainty was added quadratically to each spectral bin.
The photo-electric absorption cross sections of Morrison \& McCammon
(\cite{m:83}), ${\rm \sigma _{MM}}$, are used throughout. 
All spectral uncertainties and upper-limits are given at 90\%
confidence. Quoted count rates are corrected for
instrumental deadtime, whereas lightcurves are not.

\xmm\ observed the region of sky containing \src\ as a Target
of Opportunity (TOO) between 2000 March 21 23:55 and March 22 07:16 UTC. 
Unfortunately, the observation was curtailed for technical reasons. 
In addition, much of the data that were obtained was not correctly 
transmitted to the \xmm\ data center at Vilspa. Currently, these data
have not been recovered. An examination of the 6.7~ks of PN exposure that
is available shows that \src\ is clearly detected undergoing dipping, an
eclipse and an interval of ``persistent'' emission. However, an 
examination
of the PN spectrum shows that the instrument configuration was abnormal
and we await an updated calibration before performing a 
spectral analysis of this
data. Thus, in this {\it paper} we present only results from a 
second observation in 2001 February.

\begin{figure*}
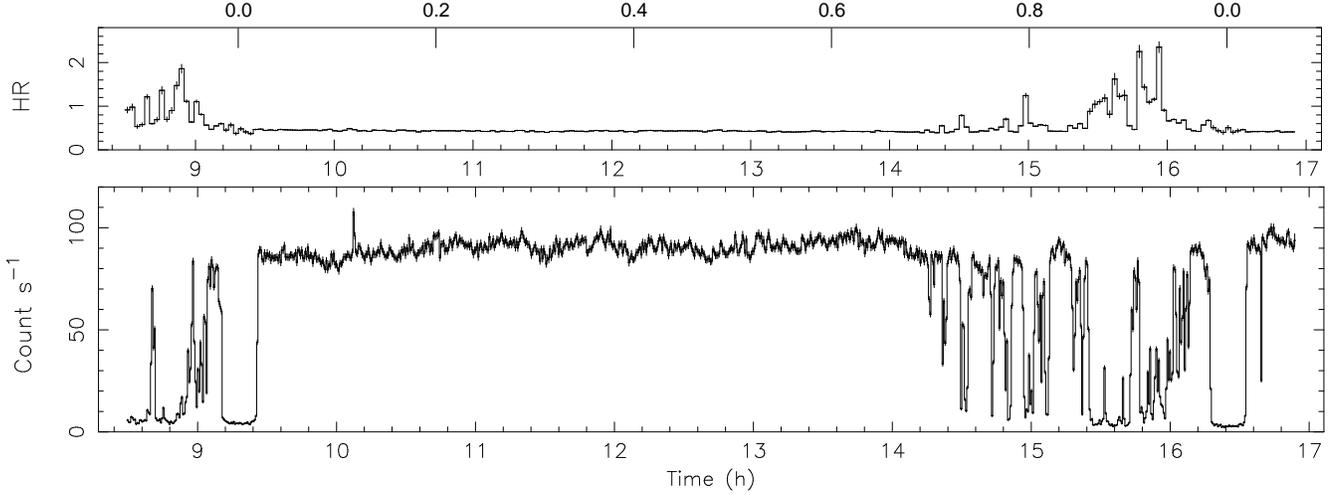

\hbox{\hspace{0.0cm}\includegraphics[width=2.48cm,angle=-90]{hr_obs2_128.ps}}
\vspace{-0.0cm}
\hbox{\hspace{0.0cm}\includegraphics[width=4.074cm,angle=-90]{obs2_lc.ps}}
  \caption[]{The \xmm\ PN 0.5--10~keV light curve of \src\ with
  a binning of 32~s. 
  Time is hours of 2001 February 20. Two eclipses, periods of intense
  dipping activity and a burst at approximately 10.1~hrs are visible. 
  The upper panel
  shows the PN hardness ratio, HR, (counts in the 2.5--10~keV energy
  range divided by those between 0.5--2.5~keV) with a binning of 128~s.
  Orbital phase is indicated above}
  \label{fig:obs2_lc}
\end{figure*}

Due to the problems encountered in the first observation, the region on
sky containing \src\ was re-observed by \xmm\ 
between 2001 February 20 08:28 and 
16:52 UTC. Table~\ref{tab:log} provides details of the 
instrument modes used. The thin filter was used with the EPIC CCDs.
The mean 2--10~keV flux obtained from the PN corresponds to a luminosity
of $1.6 \, 10^{37}$~erg~s$^{-1}$ for a distance of 15~kpc (Cominsky
\cite{c:81}).

\section{EPIC Results}

\subsection{X-ray lightcurve}

The 0.5--10~keV \src\ background subtracted
light curve obtained from the PN during the 2001 February
observation is shown in Fig.~\ref{fig:obs2_lc} with a binning of
32~s. 
At the start of the observation the source was undergoing an
interval of rapid dipping
activity, prior to an eclipse between
2001 February 20 09:10 and 09:25. 
After the first egress the source displayed a long interval of 
``persistent'' emission 
punctuated by an X-ray burst at 2001 February 20 10:12~hr. 
The height of the burst is 
strongly reduced by the binning used in Fig.~\ref{fig:obs2_lc}.
Fitting the burst spectrum 
with a blackbody, after the subtraction of  
the contribution from the persistent emission before the burst, gives
a temperature of $1.4 ^{+0.2}_{-0.1}$~keV and 
a radius of $15 ^{+7}_{-5}$~km. 
Following an interval of $\sim$6~hr, a second eclipse is visible.
After the second eclipse there is a short, deep, dip just
prior to the end of the observation.
In the 0.5--10~keV energy range 
the first and second eclipse ingresses take 
$15.2 \pm 1.1$~s and $25.1 \pm 1.1$~s (at 68\% confidence),
respectively. 
Similarly the 
egresses takes $14.4 \pm 0.8$~s and $30.4 \pm 1.0$~s.
The dips are deep with the 2--10~keV flux often reaching the
eclipses level of emission.
The upper panel of Fig.~\ref{fig:obs2_lc}
shows the PN hardness ratio (counts in the energy range 
2.5--10~keV divided by those between 0.5--2.5~keV) with a binning of
128~s. During the dips,
which occur between orbital phases $\sim$0.7--1.0, the
source becomes harder than during the persistent emission. Such
behavior is a common feature of LMXRB dips.

\begin{figure*}
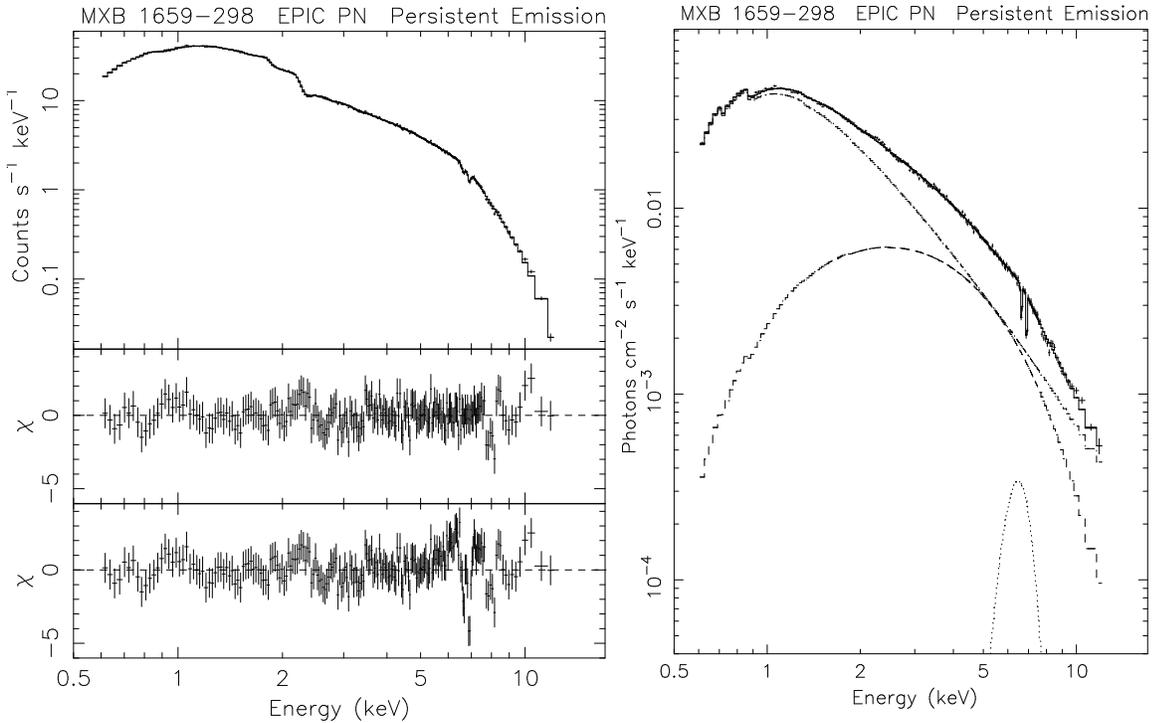

\hbox{\hspace{1.0cm}
\includegraphics[height=8.0cm,angle=-90]{f1.ps} 
\includegraphics[height=7.103cm,angle=-90]{photons.ps}}
\caption[]{The 0.6--12~keV PN persistent emission.
The left panels show the best-fit count spectrum and the 
residuals. The lowest left panel shows the residuals when the 
normalizations of the narrow absorption 
and broad emission features 
are set to zero. 
The structured residuals between 7.0--10.0~keV, as well as the 
positive residuals
around 2.2~keV are consistent with being due to instrumental effects.
The right panel shows the best-fit deconvolved photon spectrum with the 
individual components indicated. The cutoff power-law dominates 
$\approxgt$5~keV}
 \label{fig:bestfit}
\end{figure*}
 
\subsection{Persistent emission spectrum}
\label{subsect:persistent}

The persistent emission spectrum during the     
2001 February observation was investigated by
extracting the 0.2--12~keV PN events obtained
between 2001 February 20 09:30 and 14:00 UTC,
which is an interval free of eclipses and obvious dipping activity. 
Times corresponding to the X-ray burst around 
2001 February 20   10:12~hrs were
also excluded.
This selection results in a 0.6--12~keV count rate of
$125.0 \pm 0.1$~s$^{-1}$ with an exposure time of 12.0~ks.  
In the PN small window mode only a 63 x 64 pixel region
of the central CCD is read out with a time resolution of 5.7~ms
in order to reduce the effects of pile-up. 
This is expected to
be significant at count rates of $\approxgt$100~s$^{-1}$, comparable to 
that of the persistent emission (Str\"uder et al. \cite{s:01}).  
The effects of pile-up were investigated by extracting events in an
annulus outside the core of the \src\ point spread function (PSF) 
and comparing their spectrum with that obtained above. As expected,
this showed that the overall spectrum is affected by pile-up.
To minimize this, events from the inner 6\arcsec\ radius
core of the PSF were excluded from the PN spectrum which
is used in the subsequent analysis.
This leaves a background subtracted count rate of
$78.90 \pm 0.08$~s$^{-1}$.

Unfortunately, it is not yet possible to analyze MOS Timing Mode data.
The MOS2 Full Window mode data is strongly affected
by pile-up and a spectrum of the events in the wings
of the PSF was extracted. This had a count rate of
only $6.16 \pm 0.02$~s$^{-1}$. 
A comparison with the PN spectrum showed
good agreement between the two, except at energies $\approxlt$0.6~keV.
A further comparison with the RGS continuum 
(see Sect.~\ref{subsect:rgslines}) showed
good agreement between the MOS and RGS spectra and so PN data below
0.6~keV were excluded. 
Following discussion with the XMM-Newton Science Operations Center personnel,
this difference is attributed to residual 
calibration uncertainties in the PN small window mode.

We first tried to fit the 
persistent 0.6--12~keV PN spectrum of \src\ with simple models
such as an absorbed power-law, cutoff power-law 
(${\rm E^{-\alpha}\exp-(E_{c}/kT)}$), thermal  
bremsstrahlung, blackbody,
and multicolor disk-blackbody 
(Mitsuda et al.~\cite{m:84}; Makishima et al.~\cite{m:86}),
but none of these provided acceptable
fits with values of reduced $\chi ^2 >2 $ and large and structured 
residuals in all cases.
Comptonized emission models, such as the {\sc compst} and {\sc comptt}
models in {\sc xspec} (see e.g., Titarchuk~\cite{t:94}), 
were the only single-component 
models providing  reduced $\chi^2$   values of $<$2, although 
significant residuals still remained at high energies.
Indeed, the unacceptable fits can be mostly attributed to 
the presence of  
narrow negative residuals around 6.65 and 6.90~keV, suggesting
the presence of discrete absorption features.
Examination of the PN background spectra does not reveal the 
presence of features at these energies suggesting that they are not due to
improper background subtraction.

Next, we tried to fit the persistent spectrum with 
different models, always including  
the two Gaussian absorption lines at energies of 
$\sim$6.65~keV and at $\sim$6.90~keV.  
Cutoff power-law, {\sc compst} and {\sc comptt}
models were tested, giving $\chi^2$   values of 
724.2 for 253 degrees of freedom (dof),
307.3 for 253 dof and 
283.8 for 252 dof, respectively.
In all the above fits, the presence of   
broad positive residuals around 6~keV suggests the presence
of an additional Gaussian line, but this time in emission.
This reduced the $\chi^2$ to 302.5 for 250 dof, to
282.3 for 250 dof and
to 264.3 for 249 dof for the cutoff power-law,
the {\sc compst} and the {\sc comptt} continuum models, respectively.
In all three models, this additional Gaussian line is
significant at a confidence level of $>$99.99~\%.
Inspection of the residuals 
shows that, in the case of the cutoff power-law fit,
broad structured residuals $\approxlt$2~keV are still present,
suggesting the need for an additional continuum component.
Blackbody and multicolor disk-blackbody were added to the above
continuum and line models and the fits repeated.
The inclusion of these low-energy
components to the cutoff power-law and in the {\sc compst} 
models results in significantly better fits at 
a confidence level of $>$99.99~\%.
In contrast, neither low-energy component improved the fit with the 
{\sc comptt} 
model and so the best-fit with this model gives a $\chi ^2$ of
264.3 for 249 dof. 
The best-fit consists of absorbed blackbody and cutoff power-law 
continuum components
and is shown in Fig.~\ref{fig:bestfit}.
The best fit parameter values, given in Table~\ref{tab:spe},
consist of
a blackbody with temperature, kT${\rm _{bb}}$, of $1.30 ^{+0.01} _{-0.06}$~keV, 
and a cutoff power-law with a 
photon index, $\alpha$, of $1.90 \pm 0.02$ and a high energy cutoff, 
E${\rm _{c}}$,
of $15 ^{+29} _{-4}$~keV for a $\chi^2$ of 227.1 for 248 dof.
This model has been used to model the spectra
of a number of dipping LMXRB (e.g., Church et al.~\cite{c:97}), allowing an
easy comparison with results from other sources.
For comparison, if the blackbody is replaced by a disk-blackbody 
with a temperature of $2.12  ^{+0.01} _{-0.03}$~keV and a projected inner disk
radius of $2.20 ^{+0.05} _{-0.03}$~km, the $\chi ^2$ is 252.4 for 248~dof.
The best-fit \nh\ of
$(3.5 ^{+0.2} _{-0.1}) \, 10^{21}$~atom~cm$^{-2}$ is a factor $\sim$2
higher than the galactic
column of $1.8 \, 10^{21}$~atom~cm$^{-2}$ (Dickey \& Lockman \cite{d:90}),
  suggesting the presence of additional absorption.

\begin{table}
\begin{center}
\caption[]{Best-fit spectral parameters for the persistent \src\
emission with a cutoff power-law, with photon
index $\alpha$ and cutoff energy $E_{{\rm c}}$,
and a blackbody, with temperature 
$kT_{{\rm bb}}$ and radius
$R_{{\rm bb}}$ together with
2 Gaussian features in absorption and one in emission.
$E_{{\rm line}}$ is the line centroid, $\sigma$ its
width, EW the equivalent width, and $I_{{\rm line}}$ the intensity. 
Uncertainties are given at 90\% confidence for one
interesting parameter,   except for upper-limits which are 
68\% confidence. The blackbody radius and normalization 
have been corrected for the missing flux when 
extraction annuli were used and assume a distance of 15~kpc}
\begin{tabular}{llc}
\hline
\noalign {\smallskip}
Component&Parameter & Value \\
&          &         	\\
\hline
\noalign {\smallskip}
&\nh\ $(10^{22}$ atom cm$^{-2}$) &  $0.35^{+0.02} _{-0.01}$  \\
Cutoff& $\alpha$   	    &  $1.90 \pm {0.02}$   \\
power-law&$E_{{\rm c}}$ (keV) & $15 ^{+29} _{-4}$    \\
&Norm (ph.~keV$^{-1}$~cm$^{-2}$~s$^{-1}$) & $0.168 ^{+0.006} _{-0.005}$    \\
Blackbody&$kT_{{\rm bb}}$ (keV)&  $1.30^{+0.01} _{-0.06}$ \\
&$R_{{\rm bb}}$ (km) &   $4.8^{+0.07} _{-0.03}$  \\
Fe emission &$E_{{\rm line}}$ (keV)	& $6.47 ^{+0.18} _{-0.14}$ \\
feature&$\sigma$ (keV)	& $0.60 ^{+0.12} _{-0.16}$  \\
&$I_{{\rm line}}$ ($10^{-4}$~ph cm$^{-2}$ s$^{-1}$)&  $5.6 ^{+1.9} _{-1.6}$  \\
&EW 		 (eV)		&  $160 ^{+60} _{-40}$  \\
Fe~{\sc xxv}& $E_{{\rm line}}$ (keV)		& $6.64 \pm{0.02}$ \\
absorption&$\sigma$ (keV)			&   $<$0.063    \\
feature&$I_{{\rm line}}$ ($10^{-4}$~ph cm$^{-2}$ s$^{-1}$)&  $-1.2 ^{+0.3} _{-0.8}$ \\
&EW 		 (eV)		& $-33 ^{+9} _{-20}$ \\
Fe~{\sc xxvi}&$E_{{\rm line}}$ (keV)		&  $6.90 ^{+0.02} _{-0.01}$ \\
absorption&$\sigma$ (keV)			&    $<$0.043    \\
feature&$I_{{\rm line}}$ ($10^{-4}$~ph cm$^{-2}$ s$^{-1}$)&  $-1.4 ^{+0.3} _{-0.4}$ \\
&EW 		 (eV)		&$-42 ^{+8} _{-13}$ \\

&$\chi ^2$/dof                    &  227.1/248   	\\
\noalign {\smallskip}                       
\hline
\label{tab:spe}
\end{tabular}
\end{center}
\end{table}

\subsection{Line spectral features}
\label{subsect:absorption}

The two narrow absorption features found in the persistent emission PN 
spectrum 
have energies of $6.64 \pm 0.02$~keV and $6.90 ^{+0.02} _{-0.01}$~keV,
similar to the energies of the Fe\,{\sc xxv} and Fe\,{\sc xxvi}  
1s-2p transitions which occur at 6.70~keV and 6.97~keV.
These correspond to the He- and H-like ionization states of Fe
and are present over a wide range of temperatures and values of the
ionization parameter, $\xi \, (\equiv \, L{\rm /n_e \, r^{2}}$, 
where $L$ is the ionizing luminosity,
$n{\rm _e}$ is the electron density and r is the distance 
from the source). 
The measured offset of $\sim$60~eV is consistent with the
current energy calibration uncertainty of the PN small window
mode (F.~Haberl, private communication). 
When the instrumental broadening is removed the
upper-limits to the line widths correspond to velocities
of $<$${\rm 3600~km~s^{-1}}$ and 
$<$${\rm 2300~km~s^{-1}}$ for the Fe\,{\sc xxv} and 
Fe\,{\sc xxvi} transitions, respectively.
We have searched for absorption edges due to Fe~{\sc xxv} and
Fe~{\sc xxvi} at 8.83 and 9.28~keV which might be expected due to
the presence of absorption lines from these ions.
However, there is no evidence for such features, but we caution that this
is a spectral region where there is significant uncertainty in the PN
calibration.

\begin{figure}
\includegraphics[height=12.0cm,angle=0]{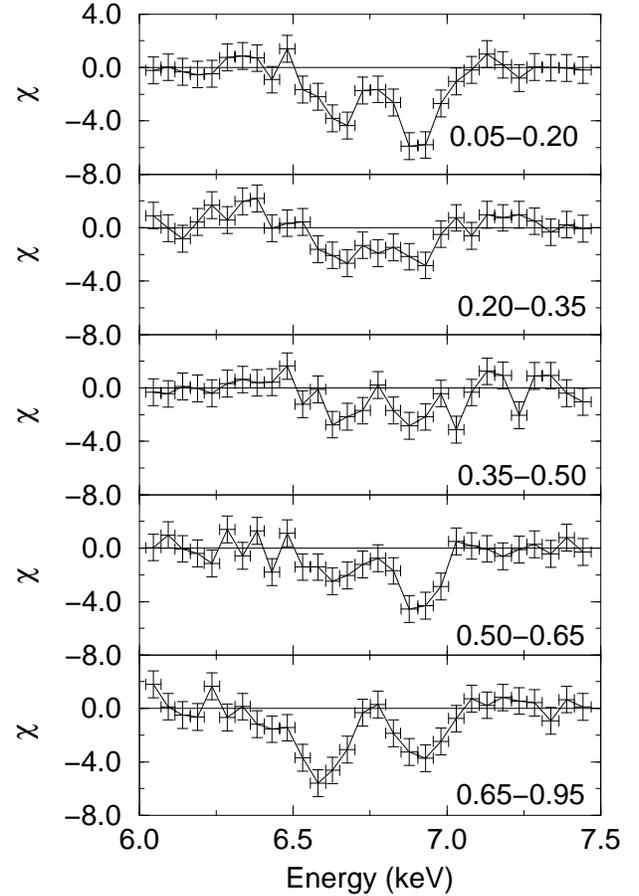}
\caption[]{Fit residuals with respect to the average
persistent emission best-fit model (see Table~\ref{tab:spe})
when the normalizations of the two narrow Fe absorption
features seen in the PN are set to zero for different 
orbital phase (indicated)}
  \label{fig:ironlines}
\end{figure}

\begin{figure}
\hbox{\hspace{-0.0cm}
\includegraphics[width=6.0cm,angle=-90]{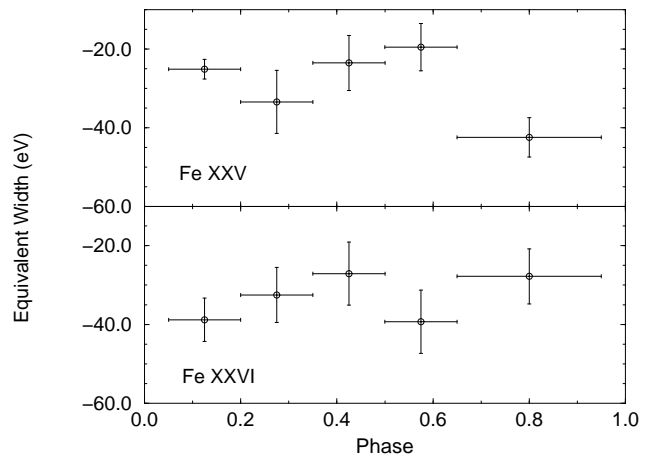}}
\caption[]{The EWs of the two Fe absorption features
as a function of orbital phase. There is no obvious orbital dependence,
even during the dipping activity at phase $\sim$0.8}
\label{fig:ew}
\end{figure}

In order to search for any orbital dependence of the absorption features, 
five PN spectra covering orbital phases between 0.05 and
0.96 (where center of eclipse is phase 0.0) were extracted.
These cover intervals of persistent emission and dipping activity
(phases 0.65--0.90).
The spectra were processed in the same way
as the persistent emission spectrum discussed 
in Sect.~\ref{subsect:persistent}.
The residuals to the persistent emission best-fit (excluding
the   line features), are shown in  
Fig.~\ref{fig:ironlines}. 
The depths of the features appear to be
variable, but there is no obvious dependence
of orbital phase (see Fig.~\ref{fig:ew}), including during the
interval of dipping activity. 
This means that similar amounts of the
material responsible for this obscuration
is present in the line of sight 
throughout most, or all, of the orbital cycle. 

A similar analysis is not possible for the broad emission feature
at 6.47~keV due to its faintness. Assuming that the same energy offset
(60~eV) should be applied to the emission feature as with the
absorption features, gives an energy of 
$6.53  ^{+0.18} _{-0.14}$~keV. Thus, the energy
is consistent with K$\alpha$ emission from a wide range of ionization
states between neutral and He-like Fe. Given the broad line width
($\sigma = 0.60 ^{+0.12} _{-0.16}$~keV) it is likely
that this feature originates from fluorescent excitation
of Fe with a range of ionization states. 

Oosterbroek et al.~(\cite{t:01}) report the possible detection
of a narrow emission line at $1.06 \pm 0.08$~keV with an EW of 
$26 \pm 16$~eV in the BeppoSAX spectrum of \src\ which they attribute
to emission from Fe-L or Ne-K ions. A similar feature
may be present at 94\% confidence in the EPIC spectrum and is visible
in the residuals of Fig.~\ref{fig:bestfit}. Adding such a narrow
feature to the best-fit model given in Table~\ref{tab:spe}
gives an energy of $0.99 \pm 0.06$~keV and an EW of $12 ^{+6} _{-8}$~eV.

\subsection{Eclipse spectrum}

As Fig.~\ref{fig:obs2_lc} indicates, there is significant flux remaining
during the eclipses. 
Counts extracted from both the eclipses give a
background subtracted PN rate of $5.29  \pm 0.09$~s$^{-1}$ for an
exposure of 0.98~ks. 
With this low count rate it is not necessary
to exclude the core of the PSF, as for the persistent emission.    

\begin{figure}
\centerline{\psfig{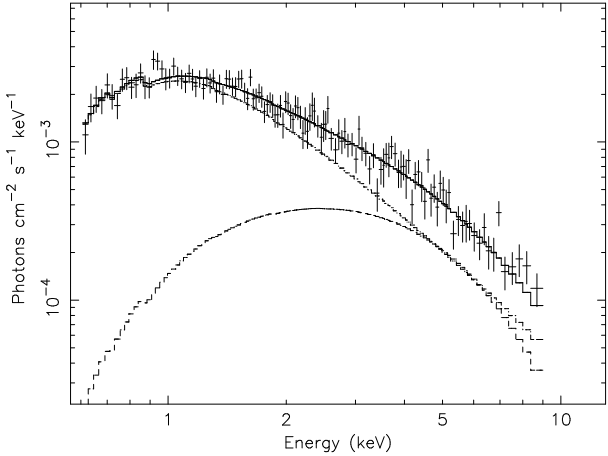}}
\caption[]{The PN eclipse spectrum of \src\ and the best-fit using the 
persistent emission spectrum consisting of a cutoff power-law
and a blackbody. The spectral parameters were fixed at the values
given in Table~\ref{tab:spe}, except for the normalizations}
   \label{fig:eclipse}
\end{figure}

The eclipse spectrum can be well fit by the same continuum
model as for the
persistent emission with all the parameters fixed at the best-fit
values given in Table~\ref{tab:spe}, 
except for the normalizations of the
two continuum components to give a $\chi ^2$ of
95.6 for 125 dof (Fig.~\ref{fig:eclipse}). 
This
model gives a 2--10~keV   absorption corrected flux
of $2.3 \, 10^{-11}$~erg~cm$^{-2}$~s$^{-1}$ which
corresponds to a 2--10~keV luminosity of $5.8 \, 10^{35}$~erg~s$^{-1}$
at a distance of 15~kpc. 
This is 3.6\% of the persistent emission.
The 90\% confidence upper-limits
to the EWs of the features at 6.47~keV, 6.64~keV
and 6.90~keV    are 380~eV,
$-121$~eV and $-42$~eV, respectively, consistent with the persistent values.

\subsection{Dip spectrum}

In order to study the \src\ spectrum during dipping intervals,
3 intensity selected spectra were extracted from intervals
that excluded persistent emission and eclipses. 
Intervals corresponding to PN 0.5--10~keV count rates 
(see Fig~\ref{fig:obs2_lc}) of $<$10~s$^{-1}$
(``deep dips''), 10--40~s$^{-1}$ (``medium dips'') and 
40--70~s$^{-1}$ (``shallow dips'') were extracted,
with a time binning of 32~s. The resulting
background subtracted PN count rates are $7.21 \pm 0.09$~s$^{-1}$,
$14.7 \pm 0.1$~s$^{-1}$, and
$50.3 \pm 0.3$~s$^{-1}$ with exposure times of 1.0~ks, 1.5~ks
and 0.5~ks, respectively. 
The core of the PSF was only excluded from the medium and shallow
dip spectra,
since the low count rate allowed the use of the entire PSF during
the deep dips without significant pile-up being present.

In order to model the spectral changes during the \src\ dips,
we used the ``complex continuum'' approach of Church et al. (\cite{c:97}) 
which has been used to model the spectral changes observed 
from a number of dipping LMXRB. Following Church et al. (\cite{c:97}), 
the shapes and normalizations of both continuum components were held
fixed at their persistent emission values.
In addition to a galactic column, $N{\rm _{gal}}$, which was fixed at
$1.8 \, 10^{21}$~\hcm\ (Dickey \& Lockman \cite{d:90}), the cutoff 
power-law was obscured by an additional partial covering component
during the dips. The three line features seen in the persistent emission
by the PN were also included in the model.
This gave unacceptable fits with $\chi ^2$   values of
310.3 for 204 dof (shallow dips), 624.7 for 226 dof (medium dips)
and 564.7 for 191 dof (deep dips). Inspection of the residuals shows
that a component with a shape similar to that of the
blackbody is still visible at low-energies, so partial covering
was added to this component as well. The intensity during the
dips is then modeled as:

$${\rm 
e^{-\sigma_{MM} {\it N}_{gal}} 
\;[[{\it f}_{bb}e^{-\sigma_{MM} {\it N}_{H_{bb}}} \rm  +\; 
(1 -\; {\it f}_{bb})] \; {\it I}_{BB} +\; }$$
$${\rm 
\; [{\it f}_{cpl}\,e^{-\sigma_{MM} {\it N}_{H_{cpl}}} \rm  +\; 
(1 -\; {\it f}_{cpl})] {\it I}_{CPL}   
+\;{\it I}_{G_1} +\; {\it I}_{G_2} +\; {\it I}_{G_3}],} 
$$
where $N_{\rm H_{bb}}$ and $N_{\rm H_{cpl}}$ are the additional
absorptions for the continuum spectral
components and $f{\rm _{bb}}$ and $f{\rm _{cpl}}$ are their 
covering fractions ($ 0 < f < 1$);
$I{\rm _{BB}}$, $I{\rm _{CPL}}$ and $I{\rm _{G_i}}$ are the 
normalizations of the blackbody, 
cutoff power-law and
the three Gaussian lines, respectively.
Thus, the only free parameters in the fitting are $N_{\rm H_{bb}}$,
$N_{\rm H_{cpl}}$, $f{\rm _{bb}}$ and $f{\rm _{cpl}}$.
The fits are significantly better with $\chi^2$   values 
of 276.4 for 203 dof,
241.6 for 225 dof, and 262.2 for 190 dof 
for the shallow, medium and deep dipping spectra, respectively
(Table~\ref{tab:dips}). 
The blackbody suffers significantly more absorption than the
cutoff power-law ($\sim$$1.5 \, 10^{24}$~atom~cm$^{-2}$
compared to $\sim$$5 \, 10^{23}$~atom~cm$^{-2}$ during the deepest dips),
but has a smaller covering fraction. Whilst the fit to the medium
dipping spectrum is reasonably good, the fits to the other two 
dip spectra are formally unacceptable and so the results
should be treated with caution.

Inspection of the fit residuals revealed structured positive residuals
around 6--7~keV. Better fits are obtained if a 
partial covering 
absorber, with an additional column of $N_{\rm H_{gau}}$ and partial covering
fraction $f_{\rm gau}$ affects the Fe emission.
If these parameters are set to the values derived for the
cutoff power-law, then the $\chi ^2$ values reduce significantly
in the case of the medium dipping interval.
Smaller reductions in $\chi ^2$ are 
evident for the deep and shallow dip spectra. 
This suggests that the iron line does indeed suffer extra absorption
during dipping intervals, but that we are unable to conclude whether
the absorbing material has similar properties as that responsible for
the absorption of the blackbody or cutoff power-law components.

\begin{figure}
\centerline{\psfig{figure=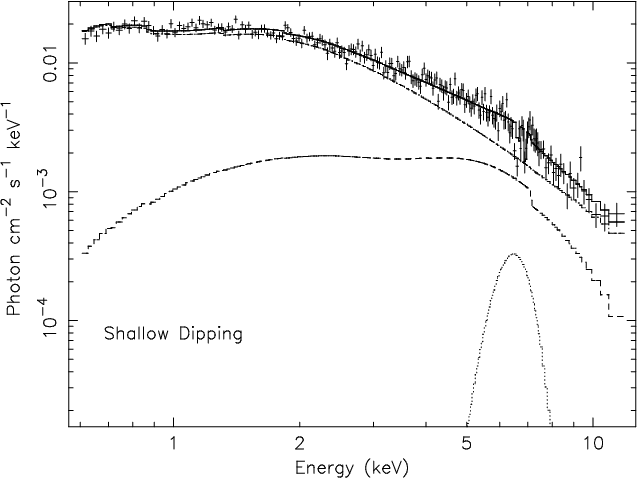,height=6.5cm,angle=-90}}
\centerline{\psfig{figure=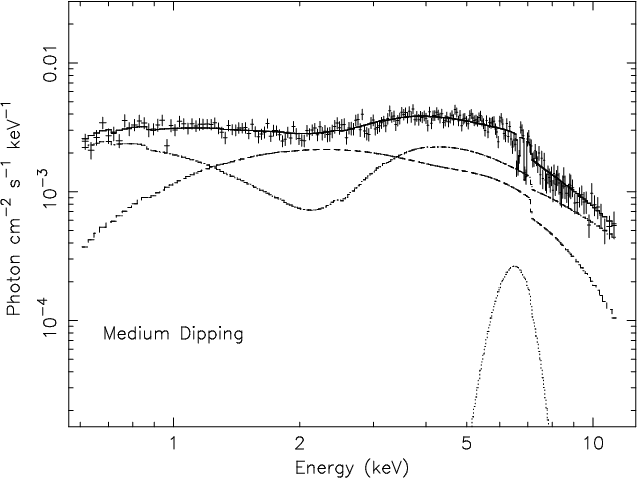,height=6.5cm,angle=-90}}
\centerline{\psfig{figure=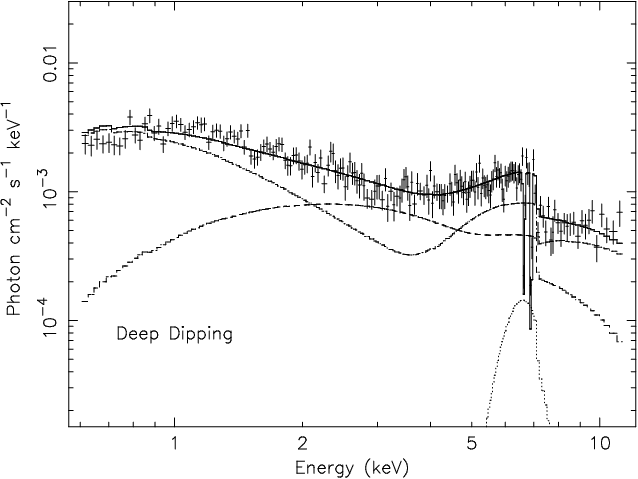,height=6.5cm,angle=-90}}
\caption[]{EPIC PN 0.6--12~keV dipping photon spectra
deconvolved using the partial covering model with the parameters
given in Table~\ref{tab:dips}. The contributions of the
cutoff power-law (which dominates at high energies), blackbody
and line features are shown separately}
   \label{fig:dips}
\end{figure}

\begin{table}
\begin{center}
\caption[]{Best-fit spectral parameters for the 3 dipping intervals.
The continuum model used is a cutoff power-law 
and a blackbody to which three Gaussian lines are added.
Only the parameters for the partial covering ({\sc pcfabs}) models
used to represent the additional absorption during dips are given. 
\nh\ is in units of $10^{22}$ atom cm$^{-2}$}
\begin{tabular}{lccc}
\hline
\noalign {\smallskip}
Parameter& \mc{3}{c}{Dip Spectrum}\\
& Shallow & Medium & Deep \\
\hline
\noalign {\smallskip}
$N_{\rm H_{bb}}$  &  $25 ^{+10} _{-8}$       & $49 ^{+7} _{-6}$            &
$157 ^{+63} _{-38}$ \\
$f_{\rm bb}$      &  $0.74 \pm 0.09$ & $0.67 \pm {0.02}$           &
$0.88\pm{0.01}$ \\
$N_{\rm H_{cpl}}$ &  $1.4 ^{+0.1} _{-0.2}$   & $14 \pm {1}$                &
$52 ^{+6} _{-3}$ \\
$f_{\rm cpl}$     &  $0.75 \pm{0.01} $       &  $0.968^{+0.002} _{-0.001}$ &
$0.960 \pm{0.002}$ \\
$\chi ^2$/dof     & 276.4/203                & 241.6/225                   &
262.2/190 \\
\noalign {\smallskip}                       
\hline
\label{tab:dips}
\end{tabular}
\end{center}
\end{table}

\section{RGS Results}
\label{sect:rgsresults}

\begin{table*}
\caption{Line features visible in the RGS persistent emission spectrum.
EW is the line equivalent width. The wavelength uncertainties do not
include a systematic uncertainty of 0.01\AA. The line widths are 1$\sigma$
upper-limits. The line widths expressed in km~s$^{-1}$ and~keV are after
removal of the instrumental broadening. Theoretical wavelengths
are from Mewe et al.~(\cite{m:85})}
\label{tab:rgslines}
\begin{tabular}{lccccccc}
\hline\noalign{\smallskip}
Ion and    & Measured         & Theoretical      & EW   & Line intensity 
&\mc{3}{c}{Line width ($\sigma$)}\\
Transition & Wavelength (\AA) & Wavelength (\AA) & (eV) & 
(ph cm$^{-2}$ s$^{-1}$) & (\AA) &(km~s$^{-1}$) & (keV)\\
\noalign{\smallskip\hrule\smallskip}

O~{\sc viii} 1s-2p & $18.95 ^{+0.02} _{-0.01}$ & 18.970 & $-2.6 \pm 0.4$ &
$-(3.9 \pm 0.6)\, 10^{-5}$ & $<$0.045 & $<$600 & $<$20\\

O~{\sc viii} 1s-3p & $16.00 ^{+0.02} _{-0.01}$ & 16.010 & $-1.3 \pm 0.4$ &
$-(3.2 \pm 1.1)\, 10^{-5}$ & $<$0.070 & $<$1200 & $<$80\\

O~{\sc viii} 1s-4p & $15.21 \pm 0.01$ & 15.180 & $-1.6 \pm 0.4$ &
$-(4.5 \pm 1.1)\, 10^{-5}$ & $<$0.058 & $<$1000 & $<$60\\

Ne~{\sc x} 1s-2p & $12.15 ^{+0.02} _{-0.01}$ & 12.130 & $-2.1 \pm 0.4$ &
$-(1.0 \pm 0.2)\, 10^{-4}$ & $<$0.049 & $<$1000 & $<$70\\

\noalign{\smallskip\hrule\smallskip}
\end{tabular}
\end{table*}

\subsection{Absorption lines}
\label{subsect:rgslines}

\begin{figure*}
\centerline{\psfig{figure=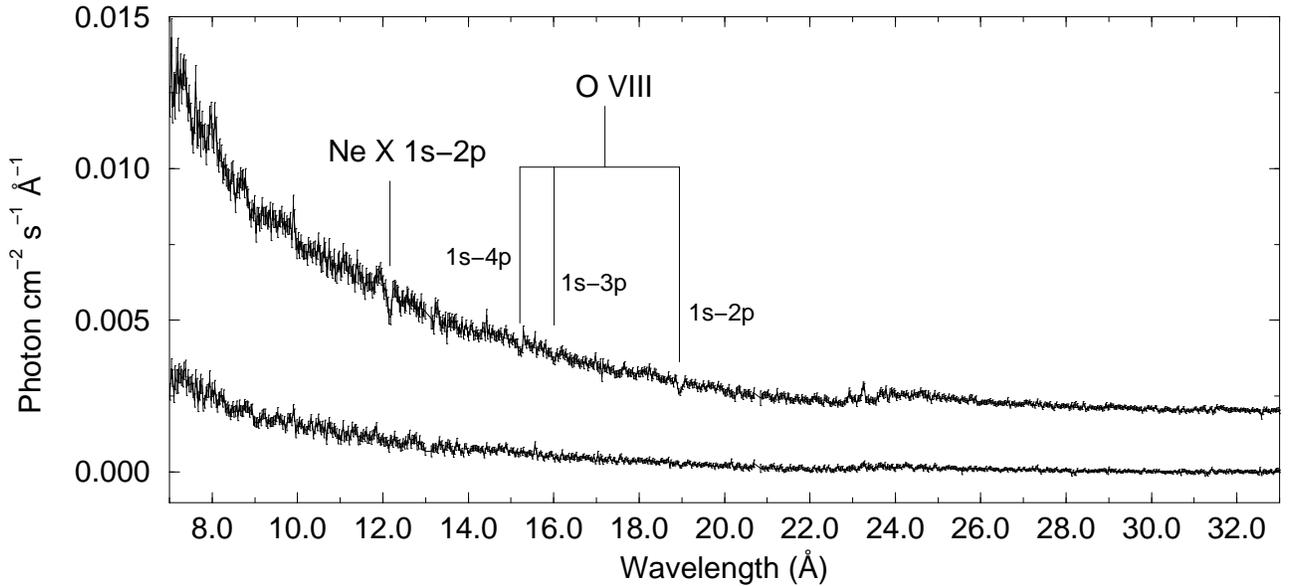,height=17.0cm,angle=-90}}
\caption[]{The first order RGS spectra of \src\ with a 
binning of 0.03~\AA. The upper curve is 
the persistent emission spectrum and the lower curve that of
the dipping intervals. Line identifications are indicated. An
offset of 0.002~photon~cm$^{-2}$~s$^{-1}$~\AA$^{-1}$ has been
added to the persistent emission spectrum for clarity
}
\label{fig:rgs12_pers_dip}
\end{figure*}

\begin{figure}
\hbox{\hspace{0.0cm}\includegraphics[height=8.5cm,angle=-90]{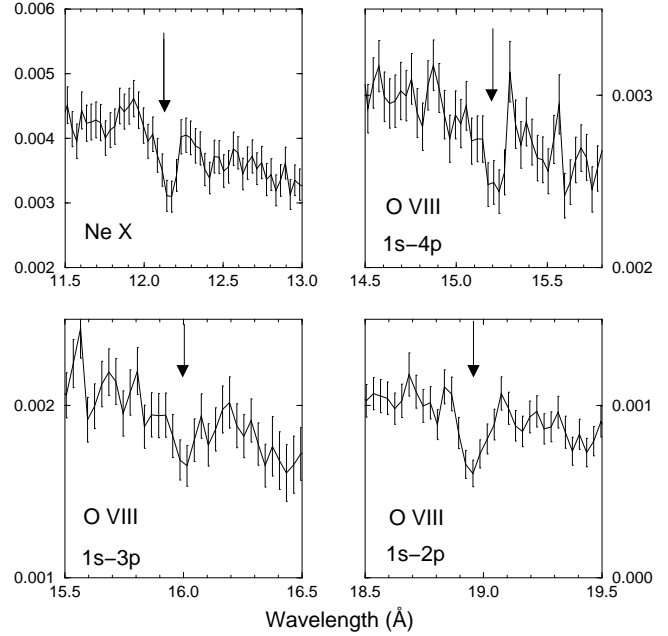}}
  \caption[]{
Persistent emission RGS spectra in the regions
around the narrow absorption features with a binning of 0.03~\AA\
in units of photon~cm$^{-2}$~s$^{-1}$~\AA$^{-1}$. 
Theoretical wavelengths are indicated}

\label{fig:rgs_spectra}
\end{figure}

The first order   RGS1 and RGS2 summed spectrum corresponding to the 
same persistent emission interval as used for the EPIC PN was extracted
and is displayed in Fig.~\ref{fig:rgs12_pers_dip}. 
The spectrum is dominated by a smooth continuum with a 
number of low-intensity features evident.
There is no evidence for the  
He-like triplet of O\,{\sc vii} or O\,{\sc viii} 1s-2p in emission
as seen from EXO\thinspace0748-676 (Cottam et al.~\cite{c:01}).
Instead, we detect 4 narrow absorption lines which we  
identify with Ne\,{\sc x} 1s-2p transition
at 12.130~\AA, O\,{\sc viii} 1s-2p at 18.970~\AA, 
O\,{\sc viii} 1s-3p at 16.010~\AA\ and 
O\,{\sc viii} 1s-4p transitions at 15.180~\AA. 
These features are shown in Figs.~\ref{fig:rgs12_pers_dip}
and~\ref{fig:rgs_spectra}
and their properties summarized in Table~\ref{tab:rgslines}.
In addition, there are unidentified features near 23\AA\ which
may be instrumental in origin.
The two 1s-2p features are clearly evident, while the
other two O~{\sc viii} transitions are less clear. Their measured
wavelengths are consistent with the theoretical values
when the uncertainty in the wavelength scale of the RGS 
(currently 0.008~\AA, den Herder et al. \cite{dh:01}) is included
(see Table~\ref{tab:rgslines}).
All the features are unresolved and their widths are
compatible with the RGS line spread function of $\sim$0.06~\AA.
When the instrumental broadening has been subtracted,
the upper-limits to the widths correspond to velocities of
$<$600~km~s$^{-1}$ and $<$1000~km~s$^{-1}$ for the O and Ne
1s-2p transitions, respectively.
Fig.~\ref{fig:rgs_spectra} shows the spectral regions around the
lines in more detail.

RGS spectra during dipping intervals (orbital phases 
0.65--0.95) were extracted and processed
in the same way as for the persistent emission spectra. 
The exposure time is 13~ks and the mean count rate 0.72~s$^{-1}$.
Fig.~\ref{fig:rgs12_pers_dip} shows the dip spectra together with those
obtained during the persistent intervals.
While the persistent spectra clearly show the two lines in absorption,
these are not so evident in the dip spectra. 
However, fitting 
narrow Gaussians to the dip spectra gives upper-limits to the EWs of 
$1.4 \pm 0.5$ and $1.5 \pm 1.0$~eV to any features at the wavelengths
of the O\,{\sc viii} and Ne\,{\sc x} 1s-2p transitions. 
Thus, we cannot exclude
the presence of these features during dipping intervals with the 
same EWs as during the persistent emission.

\section{OM Results}

A total of 20 B filter OM images were obtained.  
The counterpart to \src\ (V2134 Oph) 
is clearly recognizable in the images and is 
significantly brighter than in the $I$-band image of Wachter et
al. (\cite{w:00}). The magnitudes of V2134 Oph were obtained by
integrating the counts in a circular region with a radius of
4\arcsec\ (8 pixels).
This is slightly smaller than the recommended value of 12 pixels 
(Gondoin \cite{g:01}), but was chosen because of crowding in the
field.  The background was determined from a region 0\farcm45  
East of the source with a radius of 16 pixels.  
The background
subtracted counts were converted to magnitudes using Star~1   which
has a B-magnitude of 18.1 in the   USNO A2.0 catalog. 
Due to the relatively poor OM spatial resolution and the large 
difference in 
brightness, the contribution of the faint star close
to, and partially resolved from \src, 
is included in the magnitude determinations. 
The uncertainty in magnitude was estimated from a comparison of the 
derived Star~1 and 2 magnitudes   to be
$\sim$2\%   or 0.02 magnitudes.

\begin{figure}
\hbox{\hspace{-0.0cm}
\includegraphics[width=6.2cm,angle=-90]{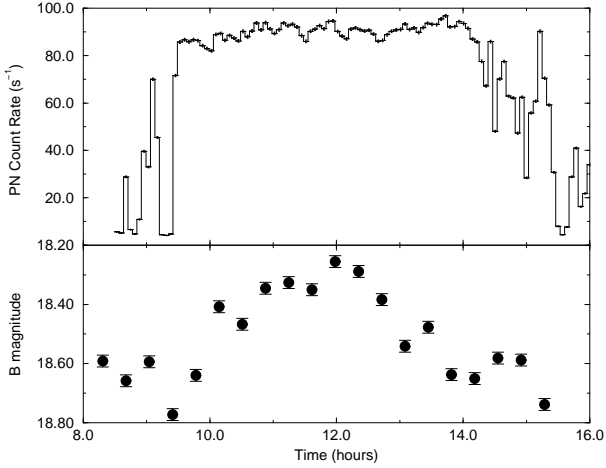}}
\caption[]{EPIC PN (0.5--10~keV) and OM ($B$ filter) lightcurves of \src.
Time is hours of 2001 February 20. An X-ray eclipse is visible at
9:20~hr}
\label{fig:ompn}
\end{figure}

Fig.~\ref{fig:ompn} shows the lightcurve together
with the rebinned 0.5--10~keV PN lightcurve. A sinusoid-like
optical modulation is present, with a narrow minimum
coincident with the X-ray eclipse superposed. 
The peak-to-peak amplitude is $\sim$0.5 magnitudes and a mean value 
$\sim$18.5~magnitudes. The overall shape of the modulation is
consistent with that reported in Wachter et al. (\cite{w:00}).
The broad modulation of the optical lightcurve with a minimum
coincident with the X-ray eclipse is consistent with the obscuration of an 
extended object, such as an accretion disk, 
centered on the compact object.
 
\section{Discussion}
\label{sect:discussion}

We report the results of an XMM-Newton observation of 
the dipping, eclipsing, bursting LMXRB \src\ in 2001 February. 
EPIC and RGS spectra of the persistent emission reveal the presence of narrow 
resonant absorption features identified with O~{\sc viii}~1s-2p, 1s-3p and 
1s-4p, Ne~{\sc x} 1s-2p, Fe~{\sc xxv}~1s-2p, and Fe~{\sc xxvi}~1s-2p 
transitions, together with a broad Fe emission feature at 
$6.47 ^{+0.18} _{-0.14}$~keV. This is in striking contrast
to EXO\thinspace0748-676 where RGS observations revealed the
presence of strong {\it emission} features from O~{\sc vii}
and O~{\sc viii} as well as weaker emission lines from
Ne~{\sc ix}, Ne~{\sc x} and N~{\sc vii} and photo-electric
absorption edges of both O~{\sc vii} and O~{\sc viii} 
(Cottam et al.~\cite{c:01}).
The EWs of the Fe absorption features observed from \src\ 
show no obvious dependence on orbital phase, 
even during dipping intervals. The EWs of the O~{\sc viii} and Ne~{\sc x}
1s-2p features are consistent with having the same values during 
persistent and dipping intervals. The absorption features are narrow
with upper-limits to any widths as low as $<$600~km~s$^{-1}$ 
(O~{\sc viii}, 1s-2p).
The overall shape of the 0.6--12~keV continuum is similar to many other
LMXRBs and may be modeled using blackbody and cutoff power-law components.
During dips the blackbody suffers significantly more absorption than the
cutoff power-law ($\sim$$1.5 \, 10^{24}$~atom~cm$^{-2}$
compared to $\sim$$5 \, 10^{23}$~atom~cm$^{-2}$ during the deepest dips).
Similar behavior has been observed from other LMXRBs
(e.g., XB\thinspace1916-053, Church et al.~\cite{c:97}),
but is in contrast to recent XMM-Newton EPIC results from
EXO\thinspace0748-676 (Bonnet-Bidaud et al.~\cite{bb:01})
and BeppoSAX results from Oosterbroek et al.~\cite{o:01}
where the hard components in these sources are seen to undergo
strong absorption during dips.  The BeppoSAX result is especially
intriguing since it is from the same source, with the same quoted
2--10~keV luminosity ($1.6 \, 10^{37}$~erg~s$^{-1}$), and
from an observation performed only 6 months earlier during the same
outburst. This change in dipping properties, if confirmed, suggests
that the location, size and structure of the {\it emitting} regions
in LMXRBs can change on a timescale of months without any apparent
change in the X-ray luminosity.   

\begin{figure}
\hbox{\hspace{-1.0cm}
\includegraphics[width=9.5cm,angle=0]{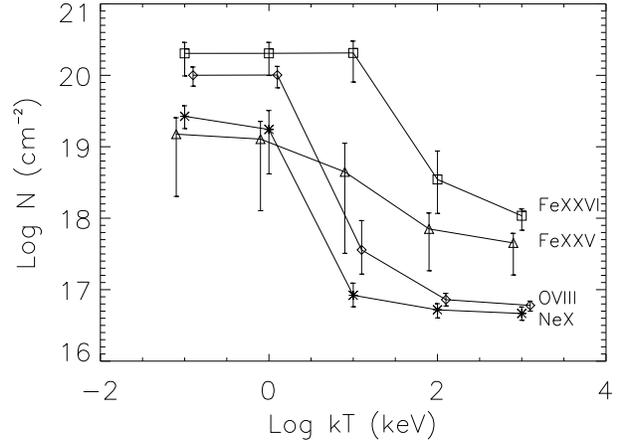}}
\caption[]{Ion column densities derived from the curve
of growth analysis for the O~{\sc viii} 1s-2p, Ne~{\sc x}, Fe~{\sc xxv} 
and Fe~{\sc xxvi} absorption lines. The curves have been 
offset slightly along the temperature axis for clarity}
\label{fig:cofg}
\end{figure}

Narrow X-ray absorption lines have been detected from the superluminal
jet sources GRO~J1655-40 (Ueda et al.~\cite{u:98}; 
Yamaoka et al.~\cite{y:01}) and GRS~1915+105 (Kotani et al.~\cite{k:00};
Lee et al.~\cite{l:01}). ASCA observations of GRO~J1655-40 revealed the
presence of absorption features due to Fe~{\sc xxv} and Fe~{\sc xxvi} which,
like those observed here, 
do not show any obvious dependence of their EWs on orbital phase.
GRO~J1655-40 has been observed to undergo deep absorption dips 
(Kuulkers et al.~\cite{k:98}) consistent with observing the source
at a high inclination of $60\degmark$--$75\degmark$ 
(e.g., Frank et al.~\cite{f:87}). ASCA observations of GRS~1915+105
revealed, in addition, absorption features due to Ca~{\sc xx},
Ni~{\sc xxvii} and Ni~{\sc xxviii}. A recent {\it Chandra} HETGS observation
of this source showed broad ($\sim$1000~km~s$^{-1}$) absorption features from
Fe~{\sc xxv} and Fe~{\sc xxvi}. 
Until now, it was possible that these absorption features are
peculiar to superluminal jet sources and related in some way to the jet
formation mechanism. With the discovery of absorption features from
the LMXRB \src\, this appears not to be the case, and as proposed
by Kotani et al.~(\cite{k:00}) ionized absorption features
may be common characteristics of accreting systems which are viewed 
close to edge-on. The discovery of these features from the two superluminal
sources may simply be a selection effect
due to their high fluxes ($\sim$1~Crab) compared to other systems.

In the likely hypothesis of photo-ionization  
(e.g., Bautista et al. \cite{b:98}), the presence of O~{\sc viii}
and Ne~{\sc x} ions is indicative of conditions where $\xi$ is
$\sim$100~erg~cm~s$^{-1}$, whereas Fe~{\sc xxv} and Fe~{\sc xxvi} are present
in regions where $\xi \approxgt 1000$~erg~cm~s$^{-1}$ 
(Kallman \& McCray \cite{kmc:82}). This broad range of $\xi$ suggests
either that the absorbing material is present over a wide range of 
distances from the central source, or has a 
large range of densities. 
The absorption features observed from \src\ are present
during a wide range of orbital phases. This excludes the possibility
that the absorbing material is associated with that responsible for the
dipping activity centered on phase $\sim$0.8.
As pointed out by Ueda et al.~(\cite{u:98}) the plasma responsible 
for the absorption features must be an-isotropically located, or
at least illuminated, around
the central X-ray source, otherwise emission from outside the line of 
sight would completely fill in the absorption features.

In order to estimate the column densities of O, Ne and Fe that produced
the observed absorption lines we performed a curve of growth analysis
following that of Kotani et al.~(\cite{k:00}) for GRS\thinspace1915+105. 
This allows the EWs of the lines to be converted into column densities
depending on the kinematic temperature of the absorbing material.
The kinematic temperature includes contributions from thermal 
motions as well as any bulk motions.
Absorbing columns for
the O~{\sc viii} 1s-2p, Ne~{\sc x}, Fe~{\sc xxv} and Fe~{\sc xxvi} features
were derived for a range of assumed temperatures (Fig.~\ref{fig:cofg}). 
As expected, a lower kinematic temperature requires a higher ion
column density, and hence a higher hydrogen column density. However
the \nh\ should not exceed one Thomson optical depth 
($1.5 \, 10^{24}$~atom~cm$^{-2}$) otherwise the absorption lines would be
strongly diminished. We assume abundances of $8.5 \, 10^{-4}$, 
$1.2 \, 10^{-4}$ and $4.7 \, 10^{-5}$ 
for O, Ne, and Fe, respectively (Anders \& Grevesse~\cite{a:89}).
Examination of Fig.~\ref{fig:cofg}
shows that the total Fe column density must be 
$\approxlt$$7 \, 10^{19}$~cm$^{-2}$ for \nh\ to be less than one Thomson
depth. This implies that $kT{\rm _{Fe} \approxgt 25}$~keV. 
A further constraint on the kinematic temperature
of the absorbing plasma can be obtained
from the upper-limits to the widths of the
lines. In the case of the Fe~{\sc xxv} and Fe{\sc xxvi} lines these
are not very constraining ($<$2400~keV and $<$1000~keV, respectively), whilst
those for the other lines, derived using Eq.~1 of Yamaoka et 
al.~(\cite{y:01}), are given in Table~\ref{tab:rgslines}.
These show that $kT{\rm _{Fe}}$ is 
inconsistent with the upper-limit
obtained from the O~{\sc viii} 1s-2p transition of $<$20~keV which
implies that the O column density is $>$$3 \, 10^{17}$~cm$^{-2}$.
This again suggests that plasma with a range of conditions is responsible
for the absorption features. Indeed, at a temperature of
25~keV $\approxgt$50\% of the Fe will be fully ionized 
(e.g., Arnaud \& Rothenflug~\cite{ar:85}) as will
virtually {\it all} of the lower $Z$ elements. This
probably explains why the source is not blanketed by the extremely
high absorption implied by the Fe lines - most of the material 
that could photo-electrically absorb at lower energies is fully ionized.
Indeed, even assuming an extremely high 
kinematic temperature of 1000~keV, the \nh\
derived from the Fe absorption features is $2 \, 10^{22}$~atom~cm$^{-2}$,
a factor of 10 higher than the excess low-energy neutral absorption
measured during quiescent intervals (Table~\ref{tab:spe}).

Following Yamaoka et al.~(\cite{y:01}) and
assuming that the plasma is in hydro-dynamical equilibrium in the
direction vertical to the disk plane, 
the distance, r, of the absorbing plasma from the central 
source can be estimated from its thermal temperature using 
$(h/r)^{2}$$\times ({\rm G} M m {\rm _H}/r$)=$kT_{\rm th}$ 
(e.g., Shakura \& Sunyaev \cite{ss:73}).
Here $M$ is the neutron star mass, G the gravitational constant, 
$m{\rm _H}$ the mass of a hydrogen atom 
and $h$ the scale height of the plasma.
Assuming that 
$h/r=\tan(90\degmark$$-i$), where $i$ is the 
binary inclination angle ($i \sim 80 \degmark$) gives 
$r_{\rm Fe} \approxlt 2.4 \, 10^{8}$~cm for 
$kT{\rm _{Fe} \approxgt 25}$~keV.
Similarly, the lower limits to $r$ derived 
from the temperatures given in Table~\ref{tab:rgslines} are
$r_{\rm O} \approxgt 3 \, 10^{8}$~cm and 
$r_{\rm Ne} \approxgt 9 \, 10^{7}$~cm. 
For a companion star mass of 0.5~${\rm \Msun}$ the binary separation 
is $2 \, 10^{11}$~cm and the radius of the accretion
disk (assumed to be $=0.8 \, \cdot R_{\rm L}$, where
$R_{\rm L}$ is the Roche Lobe radius of the neutron star) is
$5 \, 10^{10}$~cm.
Thus, the material responsible for
the Fe absorption must be located close to the central source, 
at a radius $\sim$200 times
less than that of the accretion disk, while the material responsible
for the O and Ne absorption is located at larger radii, consistent
with the lower value of the ionization parameter for these elements. 
Using the ionization parameter values ($\sim$100 for oxygen 
and neon; $>$1000 for iron), and the distance from the
central source, $r$, calculated before,
it is possible to roughly estimate the electron density, $n {\rm _e}$: 
we find $n{\rm _e}$$\sim$2$\times$10$^{18}$~cm$^{-3}$ from the O
absorption lines, 
$n {\rm _e}$$\sim$2$\times$10$^{19}$~cm$^{-3}$ from Ne and 
$n {\rm _e}$$\sim$3$\times$10$^{17}$~cm$^{-3}$ from Fe. 
Assuming that 
$N_{\rm H}$=n$_e$$\Delta r$, where $\Delta r$ is the thickness of the
absorbing layer, we derive $\Delta r \sim$200$-$4$\times$10$^{4}$~cm for
O, $\Delta r \sim$30$-$3$\times$10$^{4}$~cm for Ne 
and   $\Delta r <$5$\times$10$^{6}$~cm for Fe.

Since the absorption
features occur through a wide range of orbital phases, the absorbing material
is most likely located in a cylindrical geometry with an axis 
perpendicular to that of the accretion disk. This geometry could
explain why similar absorption features have not been so far observed
from other LMXRBs that are viewed at angles further from the orbital
plane.

\begin{acknowledgements}
The results presented are based on observations obtained 
with XMM-Newton, an ESA science mission with instruments 
and contributions 
directly financed by ESA Member States and the USA (NASA).
L.~Sidoli acknowledges an ESA Research Fellowship and thanks 
A.~Borriello, S.~Molendi and A.~Orr for helpful discussions.
We thank G.~Vacanti and F.~Haberl for assistance and T.~Kotani for making
his curve of growth software available.
\end{acknowledgements}

\end{document}